\documentstyle[jkas]{article}

\beginpage{1}
\endpage{7}
\year{2013}\volume{00}\month{January}

\runningauthor {S. TRIPPE} 
\runningtitle{GRAVITATION ON GALACTIC SCALES} 

\month{January} \year{2013} \volume{00} \issueno{00}
\beginpage{1}\endpage{7}
\date{Received ---; Accepted ---}

\begin{document}

\def\Mo{{M_{0}}}
\def\Me{{M_{\rm ex}}}
\def\Mt{{M_{\rm tot}}}
\def\ao{{a_0}}
\def\ac{{a_{\rm c}}}
\def\vc{{v_{\rm c}}}
\def\al{{\langle}}
\def\ar{{\rangle}}
\def\deg{{^{\circ}}}

%%% physics style sectioning %%%
%\def\thesection{\Roman{section}}
%\def\thesubsection{\Roman{section}.\Alph{subsection}}
%\def\thesubsubsection{\Roman{section}.\Alph{subsection}.\arabic{subsubsection}}

\makeatletter
\renewcommand*\l@section{\@dottedtocline{1}{0em}{3em}}
\renewcommand*\l@subsection{\@dottedtocline{2}{3em}{2.5em}}
\renewcommand*\l@subsubsection{\@dottedtocline{3}{5.5em}{2em}}
\makeatother

\title{A SIMPLIFIED TREATMENT OF GRAVITATIONAL INTERACTION\\ \vspace{0.5mm} ON GALACTIC SCALES}

\author{Sascha Trippe\vspace{0.5mm}}

\address{Department of Physics and Astronomy, Seoul National University, Seoul 151-742, South Korea\\ {\it E-mail : trippe@astro.snu.ac.kr}}

\address{\normalsize{\it (Received 2012 November 16; Revised 2012 December 21; Accepted 2013 January 17)}}
%\offprints{S. Trippe}

\abstract{\noindent I present a simple scheme for the treatment of gravitational interactions on galactic scales. In analogy to known mechanisms of quantum field theory, I assume \emph{ad hoc} that gravitation is mediated by virtual exchange particles -- gravitons -- with very small but non-zero masses. The resulting density and mass profiles are proportional to the mass of the gravitating body. The mass profile scales with the centripetal acceleration experienced by a test particle orbiting the central mass; this comes at the cost of postulating a universal characteristic acceleration $\ao\approx4.3\times10^{-12}$\,m\,s$^{-2}$ (or $8\pi\ao\approx1.1\times10^{-10}$\,m\,s$^{-2}$). The scheme predicts the asymptotic flattening of galactic rotation curves, the Tully-Fisher/Faber-Jackson relations, the mass discrepancy--acceleration relation of galaxies, the surface brightness--acceleration relation of galaxies, the kinematics of galaxy clusters, and ``Renzo's rule'' correctly; additional (dark) mass components are not required. Given that it is based on various ad-hoc assumptions, and given further limitations, the scheme I present is not yet a consistent theory of gravitation; rather, it is a ``toy model'' providing a convenient scaling law that simplifies the description of gravity on galactic scales.}


\maketitle

\section{INTRODUCTION \label{sect_intro}}

\noindent
It is well-known that the treatment of gravitational interaction on galactic scales comes with a serious flaw denoted as the \emph{missing mass problem}. Since the seminal works by \citet{zwicky1933} and \citet{rubin1980}, it has become clear that the \emph{dynamical} mass necessary to explain the kinematics of galactic systems exceeds the actually observed \emph{luminous} mass by up to one order of magnitude. On the one hand, this observation has lead to the postulate of non-luminous and non-baryonic \emph{dark matter} that accounts for the observed discrepancy \citep{ostriker1973}, eventually evolving into the modern $\Lambda$CDM standard model of cosmology (e.g., \citealt{bahcall1999}). On the other hand, various proposals for \emph{modified laws of inertia and/or gravity} have been made to address the missing mass problem, also in view of open questions regarding the standard cosmology (e.g. \citealt{kroupa2012}); the most successful candidate to date appears to be ``Milgrom's law'' of modified inertia \citep{milgrom1983,famaey2012}.

To date, a fully satisfactory description of gravitational interaction on galactic scales has not been found. Dark matter models are struggling with ``fine-tuning problems'' regarding the distribution of dark matter within and around galaxies (e.g. \citealt{kroupa2012}), whereas ``Milgrom's law'' is based on an -- entirely empirical -- ad-hoc modification of Newton's law of inertia (e.g. \citealt{famaey2012}).

My work aims at a simple, convenient, physically motivated description of gravity on galactic scales. In analogy to the well-established concepts of particle physics and quantum field theories (see e.g. \citealt{griffith2008} for an overview), I assume \emph{ad hoc} that gravitational interactions between masses are mediated by discrete virtual particles dubbed \emph{gravitons}. Assuming further that these gravitons have non-zero mass and obey certain rules of interaction, the resulting total mass distributions (composed of source masses plus graviton distributions) can be treated by standard Newtonian dynamics (see e.g. \citealt{binney2008} for a thorough review of stellar and galactic dynamics). The resulting scaling relations for mass profiles and rotation curves agree satisfactorily with observations.

\section{GRAVITATIONAL INTERACTION \label{sect_gravity}}

\subsection{Fundamental Assumptions \label{ssect_fundamentals}}

\noindent
In order to provide a scheme of gravitational interaction, I start with the following fundamental assumptions:

\begin{itemize}

\item[{A1.}] Gravitational interactions are mediated by discrete particles, \emph{gravitons}.

\item[{A2.}] Gravitons are \emph{virtual} particles arising from quantum fluctuations -- the mechanism commonly employed in quantum field theories.

\item[{A3.}] Gravitons have a non-zero mass.

\item[{A4.}] Interactions may only occur (1) between two real masses; (2) between a real mass and a graviton emitted by a real mass.

\end{itemize}

\subsection{Consequences \label{ssect_consequences}}

\noindent
Assumptions A1--4 lead to several consequences that, partially, can be tested by observations. Unless stated otherwise, the terms ``gravitational system'' or ``dynamical system'' denote a system composed of a point-like, real, baryonic, luminous source mass $\Mo$ and a quasi-massless test particle orbiting $\Mo$ on a circular orbit at radial distance $r$ with circular speed $\vc$.

\subsubsection{Gravitons Are ``Dark''}

\noindent
Postulating gravitons as mediators of gravitational interactions (assumption A1) implies that they couple to mass, with mass playing the role of the ``charge'' of gravity. Gravitons should not be expected to take part in electromagnetic/electroweak or strong nuclear interactions (see e.g. \citealt{griffith2008} for an overview). Accordingly, gravitons do not interact with photons electromagnetically, they are ``dark'' in the sense of being ``invisible'' to electromagnetic radiation.

\subsubsection{Graviton Masses \label{sssect_gravitonmass}}

\noindent
The creation of virtual massive particles from quantum fluctuations (assumptions A2 and A3) is, a priori, a violation of the principle of conservation of energy. This violation is only possible as long as it is ``temporary'' within the limits of \emph{Heisenberg's uncertainty relation} between uncertainties in energy and time, meaning here specifically

\begin{equation}
\label{eq_heisenberg}
\Delta E\, \Delta t \approx \hbar
\end{equation}

\noindent
where $\Delta E=m_g c^2$ is the energy of a graviton with mass $m_g$, $c$ is the speed of light, $\Delta t$ is the lifetime of the virtual particle, and $\hbar=1.055\times10^{-34}$\,Js is the reduced form of Planck's constant (e.g. \citealt{griffith2008}).

A range of gravitational interaction on cosmological scales requires graviton lifetimes on the order of the Hubble time, meaning $\Delta t \gtrsim 1.4\times 10^{10}$\,yrs. Accordingly, we require

\begin{equation}
\label{eq_gravitonmass}
m_g = \frac{\Delta E}{c^2} \lesssim 3\times10^{-69}\,{\rm kg} \approx 1.5\times10^{-33}\,{\rm eV}\,c^{-2} ~ .
\end{equation}

\noindent
This value is about 40 orders of magnitude smaller than the masses of virtual exchange particles occurring in particle physics -- a necessary consequence of the vast difference in size scales (see also \citealt{goldhaber2010} for a recent review of graviton mass limits).

\subsubsection{Density and Mass Profiles}

\noindent
I assume in the following that the source mass $\Mo$ radiates away gravitons without loosing mass itself; this is a corollary of assumption A2. Consistency with the classical theory of gravitational fields  requires that (1) the graviton density profile scales linearly with $\Mo$, and (2) the density profile obeys the inverse-square-of-distance law of gravitational fields (and of radiation fields in general). Accordingly, we find

\begin{equation}
\label{eq_density}
\rho = \Mo\,\beta\,r^{-2}
\end{equation}

\noindent
where $\rho$ is the mass density of the graviton distribution, $r$ is the radial distance from $\Mo$, and $\beta$ is a scaling parameter of the dimension of an inverse length.

The mass profile is found from integration of $\rho(r)$. Demanding that $\beta$ does not depend on $r$ explicitly,\footnote{Meaning here specifically: $\beta$ may be a function of a given well-defined distance $r'$ but not of the coordinate $r$ in general. This permits integration of $\rho(r)$ over $r$ from 0 to $r'$ and a subsequent substitution $r' \longrightarrow r$.} we find the \emph{extra mass} enclosed within a distance $r$ to be

\begin{equation}
\label{eq_massprofile}
\Me = 4\pi\,\Mo\,\beta\,r
\end{equation}

\noindent
whereas the \emph{total} enclosed mass is given by

\begin{equation}
\label{eq_totalmass}
\Mt = \Mo + \Me = \Mo\,(1 + 4\pi\,\beta\,r) ~ .
\end{equation}

\subsubsection{Enclosed Mass vs. Centripetal Acceleration}

\noindent
To permit a comparison of arbitrary gravitational systems, we need to bring the scaling parameter $\beta$ in Eq.~\ref{eq_massprofile} into a form that (1) provides a characteristic scale for a given dynamical system, and (2) does not depend on the coordinate $r$ explicitly. Both conditions are fulfilled by choosing

\begin{equation}
\label{eq_beta}
\beta = \frac{2\ao}{\vc^2}
\end{equation}

\noindent
where $\vc^2/2$ is the kinetic energy per unit mass of a test particle orbiting $\Mo$ with circular speed $\vc$, and $\ao$ is a constant of the dimension of an acceleration. Conveniently, the kinetic energy is scale-free, i.e. invariant under simultaneous scalings of coordinates and time by factors $x\neq0$. Inserting $\beta$ into Eqs.~\ref{eq_massprofile} and \ref{eq_totalmass} leads to a new expression for the enclosed total mass,

\begin{equation}
\label{eq_mass-vs-accel}
\Mt = \Mo\,\left(1 + 8\pi\frac{\ao}{\ac}\right)
\end{equation}

\noindent
with $\ac=\vc^2/r$ being the \emph{centripetal acceleration} experienced by a test particle orbiting $\Mo$ with circular speed $\vc$ at radial distance $r$ .

\subsubsection{Absence of Self-Interaction}

\noindent
Assumption A4 is crucial for the self-consistency of the ``graviton picture'' of gravitation I use. Under assumptions A1 and A3, gravity is mediated by gravitons that have mass themselves. As mass can be regarded as the ``charge'' of gravitation, we may, a priori, expect interactions also between virtual masses, leading to the creation of additional masses, that interact again, et cetera. Those interactions of higher order can result in a divergence in the creation of mediator particles, leading to massive modifications of the overall interactions -- especially to violations of the $r^{-2}$ law of gravity (and radiation in general). This is a situation commonly encountered -- and treated -- in quantum chromodynamics, expressed especially in the \emph{confinement} of quarks \citep{griffith2008}.

Assuming a quasi-massless test particle interacting with $\Mo$, assumption A4 prevents those divergences:

\begin{itemize}

\item  By permitting interactions between real masses, we ensure that the test mass ``sees'' the source mass $\Mo$ in the usual -- Newtonian -- way.

\item  By permitting interactions between real masses and virtual masses emitted by a real mass, we ensure that the test mass ``sees'' the distribution of primary gravitons emitted by $\Mo$.

\item  By preventing any other interaction, we suppress interactions of higher order, potentially leading to divergences and/or confinements.

\end{itemize}

As mentioned above, the conditions imposed by assumption A4 are far from obvious. In quantum field theories, those conditions are usually traced back to conserved quantities that impose selection rules for allowed interactions; prominent examples are spins or the color charges of quantum chromodynamics \citep{griffith2008}.

\section{TESTS \label{sect_tests}}

\subsection{Galactic Rotation Curves \label{ssect_rotcurves}}

\noindent
The circular speed of a test particle orbiting $\Mo$ is related to Eq.~\ref{eq_mass-vs-accel} like

\begin{equation}
\label{eq_velrot}
\vc^2 = \frac{G\,\Mt}{r} = \frac{G\,\Mo}{r}\left(1 + 8\pi\frac{\ao}{\ac}\right)
\end{equation}

\noindent
with $G$ denoting Newton's constant. As long as $8\pi\ao/\ac\ll1$, the first summand in the bracket dominates, resulting in (quasi-)Keplerian motion. However, in the limit $8\pi\ao/\ac\gg1$, the second summand dominates and we have

\begin{equation}
\label{eq_velrot-weaklimit}
\vc^2 \approx \frac{8\pi\,G\,\Mo}{r}\times\frac{\ao}{\ac} ~ .
\end{equation}

\noindent
Exploiting the fact that $\ac=\vc^2/r$, we can rewrite Eq.~\ref{eq_velrot-weaklimit} like

\begin{equation}
\label{eq_tullyfisher}
\vc^4 \approx 8\pi\,G\,\Mo\,\ao = const.
\end{equation}

\noindent
Accordingly, one finds that, in the limit of strong centripetal accelerations, the circular speed is Keplerian; it approaches a constant value when approaching the limit of weak centripetal accelerations. This is in agreement with the observed behavior of galactic rotation curves \citep{rubin1980}.

\subsection{Tully-Fisher/Faber-Jackson Relations \label{ssect_tully}}

\noindent
Next to the fact that rotation curves flatten out asymptotically, Eq.~\ref{eq_tullyfisher} provides additional information: it states that, asymptotically,

\begin{equation}
\label{eq_tullyfisher2}
\vc^4 \propto \Mo ~ .
\end{equation}

\noindent
This is in agreement with the empirical \emph{Tully-Fisher relation} between rotation speeds of spiral galaxies and their luminous masses \citep{tully1977}. A relation as predicted by Eq.~\ref{eq_tullyfisher} is observationally established over ten orders of magnitude of mass for gravitationally bound systems, from dwarf galaxies to galaxy clusters (cf. Fig. 48 of \citealt{famaey2012}). Likewise, Eq.~\ref{eq_tullyfisher} provides an explanation for the empirical \emph{Faber-Jackson relation} between the luminosity -- which is proportional to the source mass in my interaction scheme -- of elliptical galaxies and their velocity dispersions \citep{faber1976}.

\subsection{Mass Discrepancy--Acceleration Relation \label{ssect_mda}}

\begin{figure}[!t]
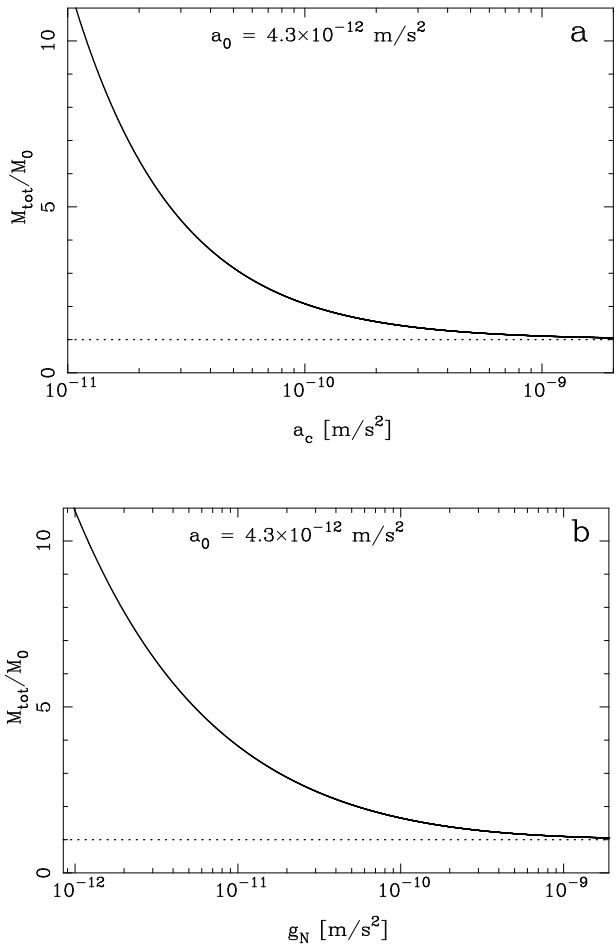

\centering
\plotfiddle{mass-accel.eps}{120mm}{-90}{35}{35}{-127}{381} \\
\plotfiddle{mass-newton.eps}{0mm}{-90}{35}{35}{-127}{214}
\caption{Theoretical mass discrepancy--acceleration relations. {\bf a.} The ratio $\Mt/\Mo$ as function of centripetal acceleration $\ac$ according to Eq.~\ref{eq_mda}. {\bf b.} $\Mt/\Mo$ as function of the Newtonian gravitational acceleration expected if $\Mo$ were the total mass, $g_N\equiv\ac/(\Mt/\Mo)$. The characteristic acceleration used here is $\ao=4.3\times10^{-12}$\,m\,s$^{-2}$. These diagrams should be compared to Fig.~10 of \citet{famaey2012}.}
\label{fig_mda}
\end{figure}

\noindent
Equation~\ref{eq_mass-vs-accel} links the total enclosed mass, the source mass, and the centripetal acceleration experienced by a test particle orbiting the source mass. By rewriting this equation slightly, we can formulate a \emph{mass discrepancy--acceleration relation} \citep{mcgaugh2004} that links the mass discrepancy, defined as the ratio $\Mt/\Mo$, and acceleration directly:

\begin{equation}
\label{eq_mda}
\frac{\Mt}{\Mo} = 1 + 8\pi\frac{\ao}{\ac} ~ .
\end{equation}

Relation \ref{eq_mda} is illustrated in Fig.~\ref{fig_mda}a for a range of weak centripetal accelerations $\ac$. In addition, Fig.~\ref{fig_mda}b shows the mass discrepancy as function of the Newtonian gravitational acceleration expected if $\Mo$ were equal to the total enclosed mass, $g_N\equiv\ac/(\Mt/\Mo)$. Comparison to Fig.~10 of \citet{famaey2012} indicates a satisfactory agreement between Eq.~\ref{eq_mda} and observations for $\ao=4.3\times10^{-12}$\,m\,s$^{-2}$ ($\pm5$\%).

\subsection{Surface Brightness vs. Acceleration \label{ssect_sigma-acc}}

\noindent
In the limit of weak centripetal accelerations, we can use Eq.~\ref{eq_tullyfisher} and rewrite it like

\begin{equation}
\label{eq_sigma}
\frac{\vc^4}{r^2} \approx 8\pi\,G\,\ao\frac{\Mo}{r^2}
\end{equation}

\noindent
For a sufficiently flat stellar system, we can define a (luminous) surface density $\Sigma\propto\Mo/r^2$. Inserting $\Sigma$ into Eq.~\ref{eq_sigma} and exploiting the fact that $\ac=\vc^2/r$, we may conclude that

\begin{equation}
\label{eq_sigma-vs-accel}
\ac^2 \propto \Sigma ~~~~~~~ {\rm or} ~~~~~~~ \ac \propto \Sigma^{1/2} ~ .
\end{equation}

\noindent
Such a correlation is indeed observed (see Fig.~9 of \citealt{famaey2012}).

\subsection{Galaxy Clusters \label{ssect_clusters}}

\noindent
The interaction scheme I use should not be restricted to galaxies but also become visible in the dynamics of galaxy clusters, likewise leading to observable mass discrepancies. Unfortunately, application of Eq.~\ref{eq_mass-vs-accel} is not straightforward because galaxy clusters are not point-like masses but extended mass distributions with most of their masses apparently stored in the X-ray luminous intra-cluster medium (e.g. \citealt{sparke2007}).

I estimate the mass discrepancy crudely -- accurate to factors of a few -- by using the velocity dispersions and core radii of the clusters as proxies for $\vc$ and orbit radii, respectively. For the four clusters Virgo, Coma, Perseus, and RDCS 1252.9--2927 (from Table 7.1 of \citealt{sparke2007}), this leads to theoretical mass discrepancies of about 2 to 4; the actually observed values are about 6 to 10. Given the crudeness of the estimate, I conclude that theoretical and observational values are consistent with each other; especially, both are of the same order of magnitude $\lesssim$10.

\subsection{``Renzo's Rule'' \label{ssect_renzo}}

\noindent
Observationally, dynamical and (luminous) baryonic surface densities in disk galaxies are closely related, expressed -- among others -- by ``Renzo's rule'' \citep{sancisi2004} which states that ``for any feature in the luminosity profile, there is a corresponding feature in the rotation curve''. This empirical rule has been used as an argument against the presence of dark matter distributions in disk galaxies because these are, a priori, independent from structures in the distributions of luminous mass \citep{famaey2012}. 

In the frame of the interaction scheme presented in this work, ``Renzo's rule'' is self-evident due to the unique relation between (luminous) source mass and total enclosed mass (Eqs.~\ref{eq_mass-vs-accel}, \ref{eq_velrot}).

\section{DISCUSSION \label{sect_discuss}}

\noindent
Starting from the ad-hoc assumptions A1--4, I arrive at a simple description of gravitational interaction on galactic scales. This ``graviton picture'' of gravity makes several predictions about galactic dynamics that are consistent with observations. Compared to standard models of galactic dynamics, the scheme I propose seems preferable because it removes the need for a separate ``dark mass'' component -- thus providing a simplified description. Notably, the ``graviton picture'' seems to be the only model to date that predicts the mass discrepancy--acceleration relation (Sect.~\ref{ssect_mda}, Fig.~\ref{fig_mda}), and thus the transitional regime between Newtonian and modified gravitational interaction, correctly -- modified laws of gravity/inertia or dark matter models in general make predictions on the \emph{asymptotic} behavior of stellar systems only.

Although it seems that my work is the first to explicitly apply the assumption of massive gravitons to galactic dynamics, the concept of massive gravitons -- or ``massive gravity'' -- itself has been discussed in field theories for more than seven decades (see, e.g., \citealt{goldhaber2010,hinterbichler2012} for reviews) and has been applied to various aspects of cosmology (e.g. \citealt{cardone2012}).

Even though my interaction scheme appears successful on galactic scales, it is obviously incomplete. It is formulated in the frame of classical Newtonian dynamics, meaning it is intrinsically non-relativistic. This implies that the scheme breaks down at small scales where relativistic deviations from Newtonian dynamics become noticeable -- on scales of the size of the solar system and less -- as well as at large scales where cosmological effects, like the Hubble flow, become evident -- on scales of mega-parsecs and beyond.

A further peculiarity of the description I provide is encoded in the formulation of the scaling parameter $\beta$ in Eq.~\ref{eq_beta}.\footnote{I also note an analogy provided by atomic physics. Rydberg's constant $R_{\infty}$ is the only constant of nature of the dimension of an inverse length (e.g. \citealt{lang2006}). In Bohr's model, $R_{\infty}$ is the inverse wavelength of a photon with the energy of the ground state of a hydrogen atom. This means that $R_{\infty}$ actually provides an energy scale; it is only a constant of nature because all hydrogen atoms are identical. Likewise, $\beta$ provides a characteristic energy scale for dynamical systems.} The choice I make is, to a certain extent, \emph{technically} self-evident: Eq.~\ref{eq_beta} provides a characteristic kinematic size scale, is independent of the radial coordinate, and is based on the scale-invariant, conserved kinetic energy. However, the apparent \emph{astrophysical} consequences discussed throughout Sect.~\ref{sect_tests} enforce a remarkable postulate: there is a characteristic acceleration $\ao\approx4.3\times10^{-12}$\,m\,s$^{-2}$ (or $8\pi\ao\approx1.1\times10^{-10}$\,m\,s$^{-2}$) which is a \emph{universal constant of nature}. The origin of this characteristic acceleration remains unclear;\footnote{Though this discussion might be loosely related to the discussion of Mach's principle; see e.g. \citet{bondi1997}.} I note however that it is well-known empirically (cf. e.g. \citealt{famaey2012}, and references therein).

Given that the interaction scheme I propose is based on various ad-hoc assumptions, and given its limitations discussed above, it is clear that the ``graviton picture'' is not yet a consistent theory of gravitation. Instead, it should be seen as a ``toy model'' that provides a simple scaling law for gravitational interaction on galactic scales and that is simpler than standard models (that require additional mass components). This seems valuable especially in view of modern numerical simulations exploring galactic to cosmological scales (e.g. \citealt{kim2011}).

Despite the various limitations stated above, one can think of several observational signatures that could be used to further probe the ``graviton picture'', in addition to the ``usual suspects'' discussed throughout Sect.~\ref{sect_tests}:

\noindent
{\bf Dynamical friction.} To a large extent, the dynamics of stellar systems is governed by dynamical friction (cf. Ch.~8.1 of \citealt{binney2008}). The assumption of dark matter halos around galaxies predicts a variety of dynamical signatures ranging from the evolution of binary star orbits (e.g. \citealt{hernandez2008}) to the effects of galaxy encounters (e.g. \citealt{dubinski1999}); accordingly, models for dark matter and/or modified laws of gravity can, and have been, examined by comparing the results of N-body simulations with observations of those systems. The ``graviton picture'' makes a strong prediction: as a corollary of assumption A4, only baryonic masses experience dynamical friction in a ``graviton halo''; the graviton distributions themselves do not interact. This is sharply distinct from the case of dark matter halos where both baryonic and dark mass particles are supposed to experience dynamical friction. Likewise, modified laws of gravity do not introduce any extra masses, meaning they do not predict dynamical friction beyond the one due to luminous mass. Accordingly, observational evidence, e.g. in galaxy collisions, for interactions corresponding to dynamical friction of baryonic mass in ``dark halos'' without interaction among those halos would provide support for the ``graviton picture''.

\noindent
{\bf Collisions of galaxy clusters.} The ``graviton picture'' demands a direct coupling of a luminous baryonic source mass and the extra mass due to the surrounding graviton distribution (cf. Eqs.~\ref{eq_massprofile}, \ref{eq_totalmass}, \ref{eq_mass-vs-accel}). In the standard dark matter picture, luminous and dark mass may be distributed independently. A test of these scenarios is provided by colliding clusters of galaxies: when the cluster cores pass through each other, the hot, X-ray bright intra-cluster gas, which comprises a large fraction of the baryonic cluster masses, experiences collisional ram-pressure and lags behind the collisionless galaxies and dark matter halos -- if those actually exist. Accordingly, a comparison of the spatial distributions of total mass -- derived from gravitational lensing of background sources -- and luminous mass is able to test dark matter and/or modified gravity models. An important example is provided by the ``Train-Wreck Cluster'' A520 \citep{markevitch2005}. In A520, observations find that the spatial distributions of total and luminous mass coincide \citep{jee2012} -- in agreement with the ``graviton picture''. Another example is provided by the ``Bullet Cluster'' 1E0657-56 \citep{barrena2002}. Here the situation is more complicated: on the one hand, observations indicate a spatial separation of dark and luminous mass; on the other hand, the cluster kinematics is found to be inconsistent with $\Lambda$CDM cosmology \citep{lee2010}.

\noindent
{\bf Gravitational lensing.} The ``graviton picture'' makes a specific prediction for the distribution of the non-luminous extra mass around a luminous source mass: as it assumes that gravity is the result of a radiation composed of massive particles, the density distribution of these particles has to obey the usual $r^{-2}$ law of radiation (cf. Eq.~\ref{eq_density}). This prediction has been tested by gravitational lensing studies that probe the density profiles of galaxies; observational results are in agreement with $\rho\propto r^{-2}$ as demanded by the ``graviton picture'' \citep{gavazzi2007}.

\noindent
{\bf Cosmology.} A non-zero mass implies a limited life time of gravitons (see Sect.~\ref{sssect_gravitonmass}). Accordingly, a decay of gravitons on cosmological time scales should modify the cosmic expansion history; especially, an acceleration corresponding to the one commonly ascribed to dark energy \citep{bahcall1999} might occur. Recent studies \citep{cardone2012} indeed conclude that massive gravity is able to reproduce the observed cosmological parameters.

Evidently, further theoretical as well as observational work is required to understand the limitations and implications of the interaction scheme I present in this study. Eventually, the ``graviton picture'' might play a role for galactic dynamics which is analogous to the one played by Bohr's ``planetary system'' model of the hydrogen atom for atomic physics: even though it is an over-simplified, coarse approximation of the actual physical situation, it provides a successful and convenient quantitative description for a wide range of phenomena. Ironically, the ``graviton picture'' resolves the ambiguity ``dark matter or modified gravity'' by assuming ``dark matter \emph{through} modified gravity''.

\section{CONCLUSIONS \label{sect_conclude}}

\noindent
Based on a set of four ad-hoc assumptions, I present a simple scheme for the treatment of gravitational interaction on galactic scales:

\begin{enumerate}

\item  I assume that gravitation is mediated by \emph{gravitons}, which are virtual, discrete exchange particles with non-zero mass. These gravitons ought to interact gravitationally but not electromagnetically, meaning they are ``invisible'' to light.

\item  Demanding consistency with classical field theory, the mass and density profiles of the graviton distribution around a source mass are proportional to the mass of the source. The density profile obeys the usual inverse-square-of-distance law of gravitational fields.

\item  The mass profile contains a scaling factor $\beta$ that can be written as the ratio of a universal, constant acceleration and the kinetic energy per unit mass of a test particle orbiting the source mass. Empirically, the characteristic acceleration is $\ao\approx4.3\times10^{-12}$\,m\,s$^{-2}$. In the chosen formulation, the mass profile is a function of the centripetal acceleration experienced by a test particle orbiting the source mass. 

\item  Mass profiles and resulting rotation curves predict the asymptotic flattening of galactic rotation curves, the Tully-Fisher/Faber-Jackson relations, the mass discrepancy--acceleration relation of galaxies, the surface brightness--acceleration relation of galaxies, the kinematics of galaxy clusters, and ``Renzo's rule'' correctly. Additional future observational tests could be provided by dynamical friction in stellar systems, collisions of galaxy clusters, gravitational lensing by galaxies, and the evolution of cosmic expansion.

\item  The interaction scheme is non-relativistic, meaning it should break down on small -- solar system  -- as well as cosmological scales.

\end{enumerate}

Within its known limitations, the ``graviton picture'' provides a useful tool for studies of galactic dynamics. Obviously, its range of validity needs to be explored more carefully in future observational as well as theoretical/numerical studies.

\acknowledgments{\noindent\small I am grateful to {\sc Junghwan Oh}, {\sc Taeseok Lee}, {\sc Jae-Young Kim}, and {\sc Jong-Ho Park} (all at SNU) for valuable technical support. This work made use of the software package {\sc dpuser}\footnote{\tt http://www.mpe.mpg.de/$\sim$ott/dpuser/dpuser.html} developed and maintained by {\sc Thomas Ott} at MPE Garching. I acknowledge financial support from the Korean National Research Foundation (NRF) via Basic Research Grant 2012R1A1A2041387. Last but not least, I am grateful to an anonymous reviewer whose careful report helped to improve this paper.}

\end{document}